\documentclass[12pt]{iopart}
\usepackage[xdvi]{graphicx}%

\newcommand{\gequ}{g_{\rm eq}}
\newcommand{\Ts}{T_{\rm s}}
\newcommand{\Tc}{T_{\rm c}}

\begin{document}

\title[An order parameter equation for the dynamic yield stress]
{An order parameter equation for the dynamic yield stress 
in dense colloidal suspensions}
\author{Michio Otsuki and Shin-ichi Sasa}

\address
{Department of Pure and Applied Sciences,  
University of Tokyo, Komaba, Tokyo 153-8902, Japan}

\ead{\mailto{otsuki@jiro.c.u-tokyo.ac.jp}, \mailto{sasa@jiro.c.u-tokyo.ac.jp}}

\begin{abstract}
We study the dynamic yield stress in dense colloidal suspensions 
by analyzing the time evolution of the pair distribution function 
for colloidal particles interacting through
a Lennard-Jones potential. We find that the equilibrium 
pair distribution function 
is unstable with respect to a certain anisotropic perturbation 
in the regime of low temperature 
and high density. By applying a bifurcation analysis 
to a system near the 
critical state at which the stability changes, we derive an amplitude 
equation for the critical mode. 
This equation is analogous to order parameter equations
used to describe phase transitions.
It is found that this amplitude equation describes  
the appearance of the dynamic yield stress,
and it gives a value of $2/3$ for
the shear thinning exponent. 
This value is related to $\delta$ in the Ising model.
\end{abstract}

\maketitle

\section{Introduction} 

%
%

Soft glassy materials, such as dense colloidal suspensions and  
super-cooled liquids, exhibit diverse rheological phenomena 
\cite{colloid,supercooled}. Typical examples are the decrease of
the viscosity with increasing the shear rate, called 
{\it shear thinning}, and the increase of the viscosity 
with the shear rate, called {\it shear thickening}.
These phenomena appear when the shear 
rate is  beyond the Newtonian regime,
that is, the regime in which the shear 
stress $\sigma_{xy}$ depends linearly on the shear rate 
$\gamma$, and hence
the viscosity, $\eta = \sigma_{xy}/\gamma$, is independent of 
the shear rate. Interestingly, the Newtonian regime becomes 
narrower as the temperature decreases and the density increases,
and eventually it disappears. 
In the case that there exists no Newtonian regime,
the stress is finite in the limit $\gamma \searrow 
0$. This stress is called the {\it dynamic yield stress}. We seek to 
understand these rheological properties systematically on the basis
of a  microscopic description of the system.

%
%

Recently, the nonlinear rheology of soft glassy materials 
has been studied extensively 
through use of  molecular dynamics simulations \cite{colloidMD,Yamamoto,Berthier},
analysis of  random spin models \cite{Kurchan1,Kurchan2}, and mode 
coupling theory applied to systems under 
shear \cite{cates1,miyazaki,cates2}. 
In particular, the mode coupling theory have predicted that
the dynamic yield stress appears discontinuously at the glass 
transition point \cite{cates1} and the results have been compared with
a numerical simulation \cite{Fuchs2} and an experiment \cite{Varnik}.
On the other hand, the power law
$\eta \sim \gamma^{-2/3}$ has been observed 
for a shear thinning fluids \cite{Berthier,Kurchan1}.

%
%

Among theoretical approaches,
one standard method for microscopic study of nonlinear rheology is
based on the mode coupling theory. In this theory, the singular 
behavior of the viscosity is described by the anomalous part of 
the time correlation function,
using a generalized Green-Kubo formula \cite{Fuchs}. 
As another approach describing the singular behavior,
one could focus on the shear stress.
Here, let us 
recall that the shear stress $\sigma_{xy}$ is defined as the average of 
the $y$ component of the inter-particle force 
acting on a plane transverse
to the $x$ direction. When there exist only two-body forces 
among the particles, 
this average can be 
expressed in terms of the pair distribution function.
Therefore, in this case, the rheological properties mentioned 
above can be accounted for through an analysis of the pair distribution function.  

%
%

Employing an approach of the latter type,
in the present Letter, we first investigate the 
linear stability of the equilibrium pair distribution function.
We find that this function is  unstable 
in the regime of low temperature and high density.
Next, applying a bifurcation 
analysis to a system near a critical state
at which the stability changes, we 
derive an order parameter equation that describes the appearance of a dynamic yield stress in a simple manner.

\section{Model}

%
%

We consider a system consisting of 
$N$ spherical colloidal particles suspended in a solvent
under shear flow described by 
$\bi{v}(\bi{r}) = (\gamma y,0,0)$.
We denote the volume of the system by $V$ and the temperature by $T$.
Let $\Gamma =(\bi{r}_1,\cdots,\bi{r}_N)$ represent the particle 
positions. The time dependent distribution function  for $\Gamma$, 
$\Psi_N(\Gamma,t)$, satisfies the Smoluchowski equation \cite{colloid} 
\begin{eqnarray}
\frac{\partial \Psi_N(\Gamma,t)}{\partial t}
=   \sum_{i=1}^N  \nabla_i \cdot \left( \frac{T}{R} \nabla_i
- \frac{1}{R} \bi{F}_i(\Gamma) - \bi{v}(\bi{r}_i) \right )
  \Psi_N(\Gamma,t).
  \label{smoleq}
\end{eqnarray}
Here, the Boltzmann constant is set to unity, 
$R$ is the friction coefficient, 
and $\bi{F}_i(\Gamma)$ is the interaction force defined by 
$
\bi{F}_i(\Gamma) = - \nabla_i \sum_{j; j \neq i} V(|\bi{r}_i-\bi{r}_j|).
$
In this work, we employ 
the Lennard-Jones potential,
with the explicit form
$V(r) = 4 \epsilon \left [ \left ( r/\sigma \right )^{-12}
- \left ( r/\sigma \right )^{-6} \right ]$.
In the treatment below, $\sigma$, $\epsilon$ and $R$ are set to unity, 
and all quantities are converted to dimensionless forms.  

%
%

From the $N$ particle distribution function $\Psi_N(\Gamma,t)$, 
the pair distribution function  $g(\bi{r},t)$ is defined as
\begin{eqnarray}
g(\bi{r},t) = V^2
\int \rmd ^3\bi{r}_3 \rmd ^3\bi{r}_4 \cdots \rmd ^3\bi{r}_N \Psi_N(\Gamma,t),
\end{eqnarray}
where
$ \bi{r} \equiv \bi{r}_1-\bi{r}_2$. Using this function,  we can 
express the time-dependent shear stress as \cite{stexpress}
\begin{eqnarray}
\sigma_{xy}(t) 
= \frac{\rho ^2}{2} \int \rmd^3\bi{r} r g(\bi{r},t)\frac{x y}{r^2}
\frac{\rmd V(r)}{\rmd r},
\label{sxy}
\end{eqnarray}
where $\rho \equiv N/V$ is 
the average number density of the colloidal particles, 
$\bi{r}=(x,y,z)$, and $r=|\bi{r}|$. Here, we have ignored 
the hydrodynamic contribution to the shear stress. 

%
%

It is useful to expand $g(\bi{r},t)$ in spherical harmonics,
and we write
\begin{eqnarray}
g(\bi{r},t) & = & f_I(r,t){\rm Im}Y_{2,2}(\theta,\phi) 
+ f_R(r,t) {\rm Re}Y_{2,2}(\theta,\phi) \nonumber \\
& + &  \sum_{l \geq 0; | m | \leq l; \{l,m\} \neq \{2,\pm 2\}}
G_{l,m}(r,t) Y_{l,m}(\theta,\phi),
\label{gsphere}
\end{eqnarray}
employing the spherical coordinate system $(r,\theta,\phi)$. 
Hence, substituting equation \eref{gsphere} into equation \eref{sxy},
we obtain 
\begin{eqnarray}
\sigma_{xy}(t) & = &  \sqrt{ \frac{2 \pi }{15}}\rho^2
\int_0^{\infty} \rmd r r^3 \frac{\partial V(r)}{\partial{r}} f_I(r,t).
\label{sxy2}
\end{eqnarray}
This expression clearly indicates that 
if there is singular behavior of the shear stress $\sigma_{xy}$,
that of the pair distribution function can be observed. 
In particular, the existence of a 
dynamic yield stress implies that the pair distribution function 
in the limit $ \gamma \searrow 0$ differs from 
the equilibrium one. This observation naturally leads us to conjecture 
that the dynamic yield stress is related to
the instability of the equilibrium pair distribution function,
$\gequ(r)$.  For this reason, we carry out a linear stability analysis of $\gequ(r)$.

%
%

In order to determine the linear stability of $\gequ(r)$,
we need to study
the time evolution of the pair distribution function.
This is described by the BBGKY hierarchy,
and hence it is determined by the three-particle
distribution function $g_3(\bi{r},\bi{r'},t)$. 
In order to obtain a self-contained description,
we truncate the BBGKY hierarchy by employing the Kirkwood 
superposition approximation, 
assuming the relation 
$g_3(\bi{r},\bi{r'},t) = g(\bi{r},t)g(\bi{r'},t)g(\bi{r}-\bi{r'},t)$
\cite{BG}.
This approximation has been  used  in the calculation of the pair distribution 
function for both an equilibrium system \cite{BGnumerical} and a 
non-equilibrium system under shear \cite{superposition}.
With this approximation, the evolution equation for the pair distribution function is 
derived as
\begin{eqnarray}
\frac{\partial g(\bi{r},t)}{\partial t} = 
- \bi{\nabla} \cdot \bi{J}(\bi{r},t), 
\label{gevolve}
\end{eqnarray}
with 
\begin{eqnarray}
\bi{J}(\bi{r},t) & = &
- 2 T \bi{\nabla} g(\bi{r},t)
- 2  \bi{\nabla} V(r) g(\bi{r},t) \nonumber \\
& & - 2 \rho \left ( \int \rmd^3\bi{r'}
\bi{\nabla} V(r')g(\bi{r},t)g(\bi{r'},t)g(\bi{r}-\bi{r'},t) \right )
\nonumber \\
& & + \gamma y \frac{\rmd}{\rmd x} g(\bi{r},t).
\label{Jap}
\end{eqnarray}

\section{Linear stability analysis}

%
%
%
%

First, the equilibrium pair distribution function $\gequ(r)$ 
for this model is obtained as the isotropic solution of the 
equation $\bi{J}=0$ in equation \eref{Jap} with 
$ \gamma=0$. This equation is called  the Born-Green equation 
\cite{BG}, which has been solved numerically \cite{BGnumerical}. 
Then, writing
\begin{equation}
g(\bi{r},t) = \gequ(r) \left ( 1 + h(\bi{r},t) \right ),
\label{gperturb}
\end{equation}
we substitute equation \eref{gperturb} into equations \eref{gevolve} 
and \eref{Jap} and obtain the form
\begin{equation}
\gequ(r) \frac{\partial }{\partial t}h(\bi{r},t) = 
\mathcal{L}(h(\bi{r},t)) + \mathcal{N}(h(\bi{r},t))
+ \gamma \mathcal{G}(h(\bi{r},t)),
\label{evolvegp}
\end{equation}
where $\mathcal{L}$ is a linear operator
and $\mathcal{N}$ contains  only second and third order 
polynomials in $h(\bi{r},t)$.

%
%

Because the equilibrium pair distribution function 
$\gequ(r)$ is non-negative, it is linearly unstable 
if and only if the linear operator $\mathcal{L}$ 
has a positive eigenvalue. 
Therefore, in order to determine the linear stability of $\gequ(r)$,
we could numerically compute the eigenvalues of $\mathcal{L}$. 
However, due to the three-dimensional spatial
dependence of $h(\bi{r},t)$,
the computation of these eigenvalues is not simple.  
Therefore, to simplify the problem, we assume perturbations $h(\bi{r},t)$ of the form
$h(\bi{r},t)= \psi(r,t) {\rm Im}Y_{2,2}(\theta,\phi)$.
Such an assumption is reasonable,
because the equilibrium pair distribution function is expected to 
be unstable with respect to perturbations of the form ${\rm Im}Y_{2,2}(\theta,\phi)$ 
in the regime characterized by low 
temperature and high density. (See equations \eref{gsphere} and 
\eref{sxy2}). 
Then, using the explicit form of $\mathcal{L}$,
we can rewrite $\mathcal{L}(h(\bi{r},t))$ in equation \eref{evolvegp} as 
\begin{equation}
\mathcal{L}(h(\bi{r},t)) ={\rm Im}Y_{2,2}(\theta,\phi) \mathcal{M}(\psi(r,t)).
\end{equation}
To solve the linear stability problem,
we seek the eigenvalues of $\mathcal{M}$.

%
%

The eigenvalues of the operator $\mathcal{M}$ are computed numerically 
in the following way. 
We first restrict the spatial domain of $\psi(r,t)$ to the interval
$[0,l]$, 
with the boundary conditions $\partial\psi(r,t)/\partial r=0$ 
at $r = 0$ and $r=l$. We next approximate the linear operator 
$\mathcal{M}$ as a matrix $\bi{M}$ by using a difference 
method with  a spatial mesh size $\delta x$ \cite{difference}. 
We calculated the eigenvalues of $\bi{M}$ for many values of 
$(\rho,T)$.

%
%

\begin{figure}[htbp]
\begin{center}
\includegraphics[height=12em]{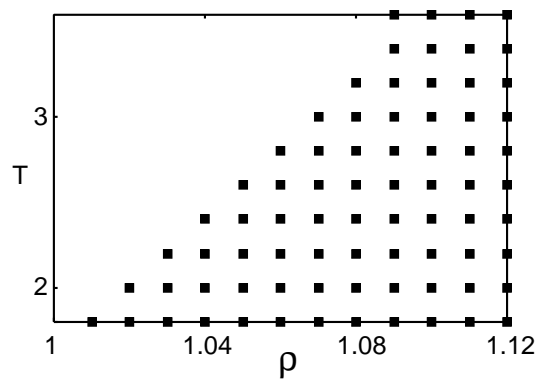}
\caption{
Stability diagram in the $(\rho,T)$ plane. The squares indicate 
states where the maximum eigenvalue of $\bi{M}$ is positive, 
and hence where $\gequ(r)$ is unstable. This result
was obtained for the numerical values $l=7.0$ and 
$\delta x = 7.0/512$. 
}
\label{phase}
\end{center}
\end{figure}

In \fref{phase}, we display the stability diagram 
obtained numerically. 
It is seen that the pair distribution 
function $\gequ(r)$ is unstable  
in the low temperature, high density regime.
Note that the form of the boundary between the two regimes is 
qualitatively similar to the curve representing the glass
transition in the $(\rho,T)$ plane calculated using mode coupling 
theory \cite{MCTmono}, though at each density, the value of the 
temperature on the boundary in the present case  is approximately 
twice that in the glass transition case. This overestimate of the 
temperature might be caused by the inaccuracy of the equilibrium pair 
distribution function $\gequ(r)$ calculated with the Kirkwood 
superposition approximation. Indeed, it has been known that the 
maximum of $\gequ(r)$ obtained by this approximation is less 
pronounced and is shifted towards smaller interparticle distances 
than that obtained by Monte Carlo simulations.

\section{Non-linear analysis}

%
%

Next, we focus on systems near the critical state at which the stability 
changes. If we fix the density, then we find that there exists a 
critical temperature $\Ts$ below which
the equilibrium pair distribution function $\gequ(r)$ is unstable. 
Let $\psi_*(r)$ be the critical 
eigenfunction of the operator ${\mathcal M}$ at $T=\Ts$. Then, we write 
a perturbation $h(\bi{r},t)$ in the form
\begin{eqnarray}
h(\bi{r},t)= A(t) \psi_*(r) {\rm Im}Y_{2,2}(\theta,\phi) + s(\bi{r},A(t)).
\label{adi}
\end{eqnarray}
Here, 
$s(\bi{r},A(t))$ represents the contribution 
to $h(\bi{r},t)$ that is not from the critical mode,
and we have assumed that its time dependence 
is restricted to that of the amplitude $A$.
This is a reasonable assumption 
because the amplitudes of non-critical modes 
decay quickly to the values determined by $A$
\cite{note:ev}.

%
%

Then, using a  bifurcation analysis (See reference \cite{Cross} as a review), 
from  equations \eref{evolvegp} and \eref{adi}, we 
perturbatively obtain the evolution equation for $A(t)$ in the form
\begin{eqnarray}
\frac{\rmd}{\rmd t} A(t) = (\Ts - T) a A(t) +G(A(t)) + \gamma H(A(t)),
\label{Geq}
\end{eqnarray}
where $a$ is a positive constant, and the functions $G(A)$ and $H(A)$
can be calculated from equation \eref{evolvegp} \cite{Otsuki2}. 
In this Letter, we do not calculate  $G(A)$ and 
$H(A)$.
Instead, considering the general forms of these functions,
we qualitatively describe the singular behavior of 
the shear stress. 

%
%

First, because the system possesses symmetry 
with respect to the simultaneous transformation 
$x \rightarrow -x$ and $\gamma \rightarrow - \gamma$, 
$G(A)$ must be odd function  and $H(A)$ an  even 
function.  We therefore expand as
\begin{equation}
G(A)=b_3 A^3+b_5 A^5+\cdots,
\label{expandG}
\end{equation}
\begin{equation}
H(A)=c_0 + c_2 A^2 + \cdots.
\label{expandH}
\end{equation}

%
%

Next, let $A_*(\gamma)$ be a stable stationary solution of 
equation \eref{Geq}. Then, from equations \eref{sxy2} and \eref{adi}, the 
shear stress for systems near the critical state $T=\Ts$ and 
$\gamma =0$ can be approximately expressed as
\begin{equation}
\sigma_{xy}  \simeq    A_*(\gamma) \sqrt{\frac{2 \pi }{15}} \rho^2
\int_0^{\infty} \rmd r r^3 \frac{\partial U(r)}{\partial{r}} 
\gequ(r)\psi_*(r),
\end{equation}
because $s(\bi{r},A(t))$ in  equation \eref{adi} 
contains no terms linear 
in $A$ as we have confirmed. Thus, the value of 
$A_*(\gamma)$ determines the behavior of the shear stress 
near the critical state. In this sense, the 
amplitude of the critical mode can be regarded as an order parameter
describing the appearance of the dynamic yield stress.

%
%

\begin{figure}[htbp]
\begin{center}
\includegraphics[height=10em]{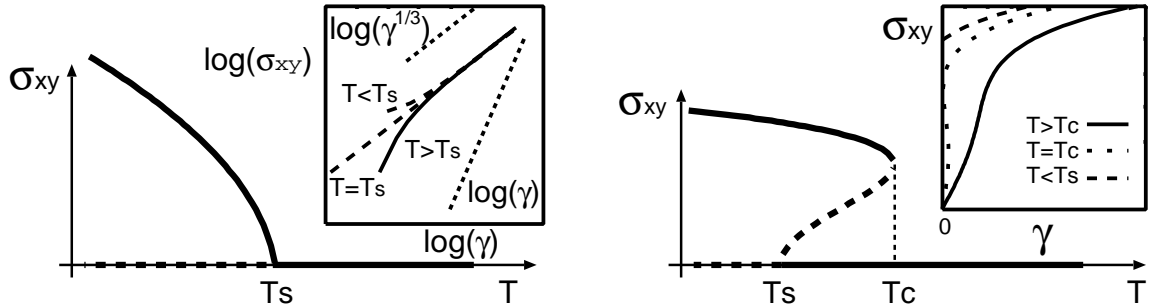}
\caption{
Shear stress in steady states 
in equation \eref{expandG}.
Here, the value plotted is that obtained
in the limit  $\gamma \searrow  0$.
The solid curve represents stable states,
and the dashed line unstable states.
The inset compares the dependence of $\sigma_{xy}$ on $ \gamma$ 
in the three cases $T>\Ts$, $T=\Ts$ and $T<\Ts$.
Left :  The case that $b_3 < 0$. Right : The case that 
$b_3 >0$ and $b_5 <0$ 
}
\label{stgra1}
\end{center}
\end{figure}

Now, we investigate equation \eref{Geq}. 
For $\gamma = 0$, this equation possesses the trivial 
stationary solution $A_*=0$. 
This solution is stable only when $ T > \Ts$. The non-trivial 
stationary solutions depend on the form of $G(A)$. First, 
let us consider the case that $b_3 < 0$ in equation \eref{expandG}. 
(We conjecture that $b_3 <0$ for the model we study.) 
Then, for $T \simeq T_{\rm s}$ and $\gamma \simeq 0$, $A_*$ satisfies 
$ (\Ts-T)aA_* +b_3A_*^3 \simeq  - \gamma c_0$ under the assumption 
that  $ A_* $ and $\gamma $ are expressed as power-law functions 
of $(T-\Ts)$.
Thus, in the limit $ \gamma \searrow  0$, the shear stress
is obtained as a function of the temperature near $T=\Ts$.
The qualitative form of this function is displayed in the left side of
 \fref{stgra1}.
It is seen that the dynamic yield stress increases continuously from 
zero as $T$ decreases from $\Ts$.
By contrast, when $T=\Ts$ and $ \gamma >0$, we obtain 
\begin{equation}
\sigma_{xy} \simeq \gamma^{1/\delta},
\label{spow}
\end{equation}
with $\delta = 3$.
This $\delta$ corresponds to that
appearing in the relation $M \simeq H^{1/\delta}$ describing critical phenomena
for magnetic materials, and the value $\delta = 3$ is the mean field value
for the Ising model. Note that equation \eref{spow} yields
$\eta \simeq \gamma^{-(1-1/\delta)}$.
The exponent $1-1/\delta$ is called the {\it shear thinning exponent},
and we find it to be $2/3$ in this analysis. We also find that 
when $T > \Ts$,  a Newtonian regime appears near $\gamma =0$. 
This regime is connected to the power law regime that exists for 
higher shear rates. 
As seen in the inset of the left side of \fref{stgra1}, this behavior 
corresponds to shear thinning. 

Next, we consider the case
$b_3 > 0$ and find that here,
a qualitatively different form of
the stress is obtained. (We expect that there do exist models
in which $b_3 >0$,
although we believe that the model studied presently is not one of
these.)
One physically plausible situation is that
in which $b_5 <0$, where we have the form displayed in the right side
of \fref{stgra1}. 
In this case, 
as the temperature is decreased, 
the dynamic yield stress increases discontinuously 
from zero to a finite value at some temperature  $\Tc$($>\Ts$),
as shown in the right side of \fref{stgra1}. Furthermore,
it is easily confirmed that shear thickening exists when $T>\Tc$.

\section{Conclusion}

%
%

The main finding of this Letter is that the order parameter equation 
given in equation \eref{Geq} provides a simple description of the nonlinear
rheological phenomena exhibited by the systems considered here.
We note that the order parameter in our model 
is defined to be the amplitude of  the critical mode of the pair 
distribution function, which is related to the equal time correlation 
function of the density.
This contrasts with the situation for the critical phenomena of 
liquid-gas phase transitions, 
for which the order parameter is defined in terms of the critical mode of the density.
Consideration of this difference might help 
elucidate the essence of the glass transition. 

We conjecture that the coefficient $b_3$ is negative for the model we
 study.
This indicates that the dynamic yield stress appears continuously from
$0$ at the transition point. This result is different from that 
obtained by the mode coupling theory that predicts the discontinuous 
appearance of the dynamic yield stress. Note that the recent result in 
Ref. \cite{Fuchs2} might support the discontinuous onset,
while the power-law behavior consistent
with our result is still observed in the range $10^{-4} < \gamma < 10^{-1}$ in this report.
Our simple theory in the present version might miss some important
physical effects. We will study further in order to understand this
discrepancy.

Finally, we present remarks regarding the two assumptions we employed 
in our analysis. 
First, the perturbations to $\gequ(r)$ are assumed to be 
restricted to the form 
$h(\bi{r},t)= \psi(r,t) {\rm Im}Y_{2,2}(\theta,\phi)$ for simplicity.
Here, we note that we already 
performed the analysis without this restriction and obtained the same 
stability diagram \cite{Otsuki2}. In this analysis, 
the five critical modes, which correspond to $\sigma_{xy}$, $\sigma_{yz}$, 
$\sigma_{zx}$,  $\sigma_{xx}-\sigma_{yy}$, 
and $\sigma_{zz}-(\sigma_{xx} +\sigma_{yy})/2$, appear 
simultaneously at $T=T_{\rm s}$ as the result of the rotational 
symmetry. Then, five order parameters are defined in association with 
these critical modes. We will report the result in another paper.

Second, with regard to the Kirkwood superposition approximation, we consider that
it has some analogy with a mean field theory for critical phenomena. 
Therefore, we expect that the time evolution of the  five order parameter 
fields (that depend on the spatial coordinate) under the influence of 
noise can describe rheological phenomena more precisely. This extension 
might modify the critical exponent $\delta$, and 
also it might play an important role in the description of phenomena 
for the case $T < T_{\rm s}$. We wish to develop such a theory as 
a natural extension of our analysis in this Letter.

%
%
\ack
This work was supported by a grant from the Ministry of 
Education, Science, Sports and Culture of Japan (No. 16540337).


\section*{References}

\begin{thebibliography}{10}


\bibitem{colloid} Russel W B, Saville D A and Schowalter W R,
{\it Colloidal Dispersions } (New York: Cambridge University Press).

\bibitem{supercooled} Larson R G, 1999
{\it The Structure and Rheology of Complex Fluids }
(New York: Oxford University Press).

\bibitem{colloidMD} Strating P, 1995 {\it J. Chem. Phys.} {\bf 103} 10226

\bibitem{Yamamoto} Yamamoto R and Onuki A, 1998 {\it Phys. Rev.} E 
{\bf 58} 3515

\bibitem{Berthier} Berthier L and Barrat J L, 2002 {\it J. Chem. Phys.}
{\bf 116} 6228


\bibitem{Kurchan1} Berthier L, Barrat J L and Kurchan J, 2000 {\it Phys. Rev.} E {\bf 61} 5464

\bibitem{Kurchan2} Sellitto M and Kurchan J, 2005 {\it Phys. Rev. Lett} {\bf 95} 236001

\bibitem{cates1} Fuchs M and Cates M E, 2002 {\it Phys. Rev. Lett.} {\bf 89} 248304

\bibitem{miyazaki} Miyazaki K, Reichman D R and Yamamoto R, 2004 {\it Phys. Rev.} E {\bf 70} 011501

\bibitem{cates2} Holmes C B, Cates M E, Fuchs M and Sollich P, 2005 {\it J. Rheol.} {\bf 49} 237

\bibitem{Fuchs2}
Fuchs M and Ballauff M, 2005 {\it J. Chem.  Phys.} {\bf 122} 094707

\bibitem{Varnik}
Varnik F and Henrich O, 2006 {\it Phys. Rev. B} {\bf 73} 174209



\bibitem{Fuchs} Fuchs M and Cates M E 2005 {\it J. Phys: Condens. Matter} {\bf 17} S1681


\bibitem{stexpress} Kirkwood J G, Buff F P and Green M S, 1949 {\it J. Chem. Phys.} {\bf 17} 988

\bibitem{BG} Green H S, 1952
{\it The Molecular Theory of Fluid} (Amsterdam: North-Holland)

\bibitem{BGnumerical} Broyles A A, 1960  {\it J. Chem. Phys.} {\bf 33} 456

\bibitem{superposition} Ohtsuki T, 1981 {Physica} {\bf 108A} 441


\bibitem{difference} Delves L M and Mohamed J L, 1985
{\it Computational Methods for Integral equations}
(London: Cambridge).

\bibitem{MCTmono} Bengtzelius U, 1986 {\it Phys. Rev. A} {\bf 33} 3433 

\bibitem{note:ev} 
We found numerically that 
in the neighborhood of the stability boundary,
the largest non-critical eigenvalue is less than $-0.5$.


\bibitem{Cross}
Cross M C and Hohenberg P C, 1993  {\it Rev. Mod. Phys.} {\bf 65}
851

\bibitem{Otsuki2}
Otsuki M and Sasa S, 2006 in preparation.


\end{thebibliography}
\end{document}